\ifdefined\pdfminorversion
\fi
\documentclass[10pt,twocolumn,letterpaper]{article}
\usepackage[T1]{fontenc}
\usepackage[utf8]{inputenc}
\usepackage{newtxtext,newtxmath}
\usepackage[margin=0.75in,columnsep=0.25in]{geometry}
\usepackage{microtype}
\usepackage{amsmath}
\usepackage{graphicx}
\usepackage{adjustbox}
\usepackage{booktabs}
\usepackage[table]{xcolor}
\usepackage{xspace}
\usepackage{listings}
\usepackage[font=small,labelfont=bf]{caption}
\usepackage[numbers,sort&compress]{natbib}
\usepackage[hidelinks]{hyperref}
\hypersetup{
  pdftitle={WISE: An Agentic Framework for Whole-page Multimodal Search Experience Attribution},
  pdfauthor={Wenyi Xu, Renzhe Zhou, Songyang Li, Feiran Zhu, Ruoqi Zhao, Chenglei Dai, Xuanping Li, Kaiqiao Zhan, Jingwei Zhuo}
}
\makeatletter
\renewcommand{\@maketitle}{\newpage
  \null
  \vskip -0.5em
  \begin{center}{\LARGE\bfseries \@title\par}\vskip 1em
    {\normalsize \lineskip 0.4em \@author\par}\end{center}\vskip 0.8em
}
\makeatother

\definecolor{promptbg}{RGB}{247,248,250}
\definecolor{promptframe}{RGB}{205,210,218}
\lstdefinestyle{prompt}{
  basicstyle=\ttfamily\scriptsize,
  backgroundcolor=\color{promptbg},
  frame=single,
  rulecolor=\color{promptframe},
  breaklines=true,
  breakatwhitespace=false,
  columns=fullflexible,
  keepspaces=true,
  showstringspaces=false,
  aboveskip=5pt,
  belowskip=5pt,
  xleftmargin=2pt,
  xrightmargin=2pt
}

\newcommand{\method}{\textsc{WISE}\xspace}
\newcommand{\bench}{\textsc{WISE-Bench}\xspace}
\newcommand{\repair}{\textsc{ReT}\xspace}
\newcommand{\best}[1]{\textbf{#1}}

\begin{document}

\title{WISE: An Agentic Framework for Whole-page Multimodal Search Experience Attribution}

\author{
  Wenyi Xu$^{1}$\thanks{These authors contributed equally to this work.}\quad
  Renzhe Zhou$^{1}$\footnotemark[1]\quad
  Songyang Li$^{1}$\footnotemark[1]\quad
  Feiran Zhu$^{2}$\quad Ruoqi Zhao$^{1}$\\[0.4em]
  Chenglei Dai$^{1}$\quad Xuanping Li$^{1}$\quad Kaiqiao Zhan$^{1}$\quad
  Jingwei Zhuo$^{2}$\thanks{Corresponding author: \texttt{zhuojw10@gmail.com}.}\\[0.7em]
  $^{1}$Kuaishou Technology\qquad
  $^{2}$Independent Researcher\\[0.5em]
  {\scriptsize\texttt{\{xuwenyi03,zhourenzhe03,lisongyang03,zhaoruoqi,daichenglei,lixuanping,zhankaiqiao\}@kuaishou.com}}\\
  {\scriptsize\texttt{a499616042@163.com}\quad\texttt{zhuojw10@gmail.com}}
}
\date{}
\maketitle

\begin{abstract}
    Algorithm-driven services, including search, recommendation, and advertising, rely on multi-stage model pipelines. While research prioritizes model optimization, post-deployment monitoring focuses on operational reliability. 
    Yet operationally healthy systems may still produce poor-quality results, exposing a gap in existing evaluation and monitoring practices. 
    However, a systematic problem formulation and benchmark for this task remain lacking.
    To bridge this gap, we formulate System Quality Diagnosis (SQD) for algorithm-driven services, a new task that extends quality evaluation beyond identifying poor outcomes to attributing their causes across system components.
    Focusing on industrial search, we propose WISE, an agentic framework for end-to-end quality diagnosis that discovers user-facing issues and traces their origins to underlying algorithmic components. 
    To improve diagnostic reliability, we introduce ReTrace for constructing evidence-consistent trajectory supervision and Adaptive Harness--Policy Optimization (AHPO) for failure-guided optimization of the agent harness and model policy. To enable systematic evaluation, we construct WISE-BENCH, a real-world multimodal search benchmark that combines user-facing evidence, internal pipeline traces, and expert diagnosis annotations.
    Extensive experiments show consistent gains across model scales, with WISE achieving 75.20\% end-to-end accuracy on Qwen3.5-35B-A3B. In production, WISE supports quality auditing, online quality assurance, and model optimization, reducing diagnosis costs and improving downstream relevance models.

\end{abstract}


\section{Introduction}

Modern algorithm-driven services, including search, recommendation, and advertising, rely on multi-stage pipelines of learned models~\cite{covington2016youtube,sculley2015debt}. While research has extensively studied model evaluation and optimization, post-deployment monitoring primarily focuses on operational reliability, such as latency, availability, and errors. Yet operationally healthy systems may still produce poor-quality outcomes. For example, a search request may successfully pass through retrieval and ranking while returning results that fail to satisfy user intent. Such failures highlight a gap between ensuring that a system operates correctly and understanding whether it delivers high-quality results, as well as where quality problems originate.

Existing studies have advanced quality evaluation and optimization for algorithm-driven services~\cite{choi2025bloomintent,ecomteam2026case}, while recent agentic RCA methods focus on diagnosing operational failures~\cite{chen2024rcacopilot,xu2025openrca}. However, a unified formulation and benchmark for diagnosing outcome-quality failures remain lacking. To bridge this gap, we formulate \textbf{S}ystem \textbf{Q}uality \textbf{D}iagnosis, \textbf{SQD} extending quality evaluation beyond identifying poor outcomes to attributing their causes across system components. Unlike conventional fault diagnosis, SQD must recognize implicit, context-dependent quality issues and investigate their origins across interacting algorithmic components. Similar quality problems may arise from different stages, while heterogeneous internal evidence may require different investigation paths. These characteristics make SQD a challenging task that demands both semantic quality understanding and evidence-grounded reasoning.

To address these challenges, we propose \textbf{WISE}, an agentic framework for end-to-end quality diagnosis in industrial search. WISE integrates multimodal quality assessment with internal evidence investigation through a structured diagnostic workflow, supported by an LLM policy and an execution harness. However, building a reliable diagnostic agent presents further challenges. Expert annotations provide diagnostic outcomes but lack the execution trajectories required for agent training. We therefore introduce ReTrace, which constructs evidence-consistent supervision through trajectory repair. Moreover, diagnostic failures may arise from either incorrect model decisions or limitations in the execution harness. To address these distinct failure modes, we propose Adaptive Harness--Policy Optimization (AHPO), which adaptively optimizes the agent policy or harness based on localized diagnostic failures.

To bridge the gap between search quality evaluation and pipeline-level diagnosis, we introduce WISE-Bench, a real-world multimodal search benchmark combining user-facing evidence, internal pipeline traces, and expert diagnostic annotations. Extensive experiments demonstrate the effectiveness of WISE across model scales and validate the contributions of ReTrace and AHPO. Furthermore, WISE has been deployed in production for quality auditing, online quality assurance, and model optimization, demonstrating its practical value in improving search quality and operational efficiency.

The main contributions of this paper are:
\begin{itemize}
    \item We formulate SQD and introduce WISE-BENCH to jointly evaluate the discovery of quality issues and the attribution of their causes to internal system components in industrial search.
    \item We develop WISE for agentic diagnosis, together with ReTrace for evidence-consistent trajectory supervision and AHPO for failure-guided harness--policy optimization.
    \item We demonstrate substantial diagnostic gains across model scales and validate the practical value of WISE in production search quality auditing, online quality assurance, and model optimization.
\end{itemize}

\section{Related Work}

\subsection{Quality Evaluation and Optimization}

Existing studies have extensively explored quality evaluation in search and recommendation systems \cite{zangerle2022evaluating}. Recent benchmarks, including M-BEIR, MMSearch-Plus, and SVFSearch, evaluate multimodal retrieval and agentic search capabilities \cite{wei2024uniir,tao2025mmsearchplus,mao2026svfsearch}. Recent methods leverage LLMs to assess user-facing quality, such as BloomIntent \cite{choi2025bloomintent}, which evaluates search results against fine-grained user intents. Beyond evaluation, agentic frameworks increasingly connect quality assessment with optimization. A case-driven multi-agent framework \cite{ecomteam2026case} automates bad-case analysis and resolution in e-commerce search. Self-EvolveRec \cite{kim2026selfevolverec} combines qualitative user feedback with internal model diagnosis to guide recommendation optimization, while SR-Agent \cite{yang2026sragent} refines post-ranking strategies based on identified experience issues. Despite these advances, existing search benchmarks primarily evaluate user-facing outcomes, lacking the internal pipeline evidence and attribution annotations needed for system-level quality diagnosis.

\subsection{Root Cause Analysis}

Root cause analysis (RCA) has been widely studied for diagnosing failures in complex software systems. Traditional methods, such as MicroRCA \cite{wu2020microrca}, localize performance anomalies through system metrics and dependency information. Recent LLM-based approaches, such as RCACopilot \cite{chen2024rcacopilot}, further automate incident diagnosis. Meanwhile, OpenRCA, OpenRCA 2.0, ORCA-bench, and AIOpsLab provide benchmarks and environments for evaluating automated fault diagnosis using heterogeneous operational evidence \cite{xu2025openrca,fang2026openrca2,gong2026orcabench,chen2025aiopslab}. However, these works primarily focus on operational failures rather than the quality of algorithmic outputs. In contrast, SQD requires identifying semantic quality issues from multimodal search results and attributing them to internal algorithmic components. Existing benchmarks therefore do not jointly evaluate user-facing quality issue discovery and evidence-grounded pipeline attribution.

\subsection{Agent Training and Optimization}
Process supervision studies feedback beyond final answers~\citep{lightman2023verify}. Search-R1 and ReTool learn search and tool-use behavior through reinforcement learning~\citep{jin2025searchr1,feng2025retool}, while Agent Lightning separates agent execution from policy training~\citep{luo2025agentlightning}. We use GRPO for policy optimization~\citep{shao2024deepseekmath}, with endpoint-supervised trajectory repair providing the initial SFT data. Unlike unrestricted regeneration, repair preserves the observations in a recorded execution and revises the supported decision. Causal Agent Replay attributes failures through intervention and re-execution~\citep{shah2026causalreplay}. Our failure analyzer instead identifies the earliest supported error from the observed trace; without counterfactual replay, this localization does not establish a causal effect.

\section{Problem and Benchmark}
\label{sec:task}

\subsection{Problem Formulation}
We formulate SQD as the task of identifying quality issues in the outputs of algorithm-driven services and attributing their causes to underlying system components using available evidence. In this work, we instantiate SQD in industrial search.
A search case is denoted by $x=(q,t,s,L)$, where $q$ is the query, $t$ is the request time, $s$ is the search scenario or source, and $L=[i_1,\ldots,i_K]$ is the displayed result list in exposure order. We distinguish two sources of evidence. Experience evidence describes the request and user-facing result content, including text, ASR, OCR, visual content, publication time, and author or source information. Pipeline evidence describes how the search system produces the displayed results, including candidate supply, recall paths, model scores, and stage-wise ranking information.
Given a search case and the available evidence, search experience diagnosis aims to produce a structured diagnosis
\begin{equation}
G(x)=(\delta,\mathcal{I},\mathcal{U},\mathcal{A},\mathcal{E}),
\end{equation}
where $\delta\in\{0,1\}$ indicates whether the case
contains an experience problem,
$\mathcal{I}\subseteq L$ is the set of problematic items,
$\mathcal{U}$ denotes the experience issues,
$\mathcal{A}$ denotes the pipeline attributions,
and $\mathcal{E}$ contains the supporting evidence.

\begin{figure*}[t]
    \centering
    \includegraphics[width=\textwidth]{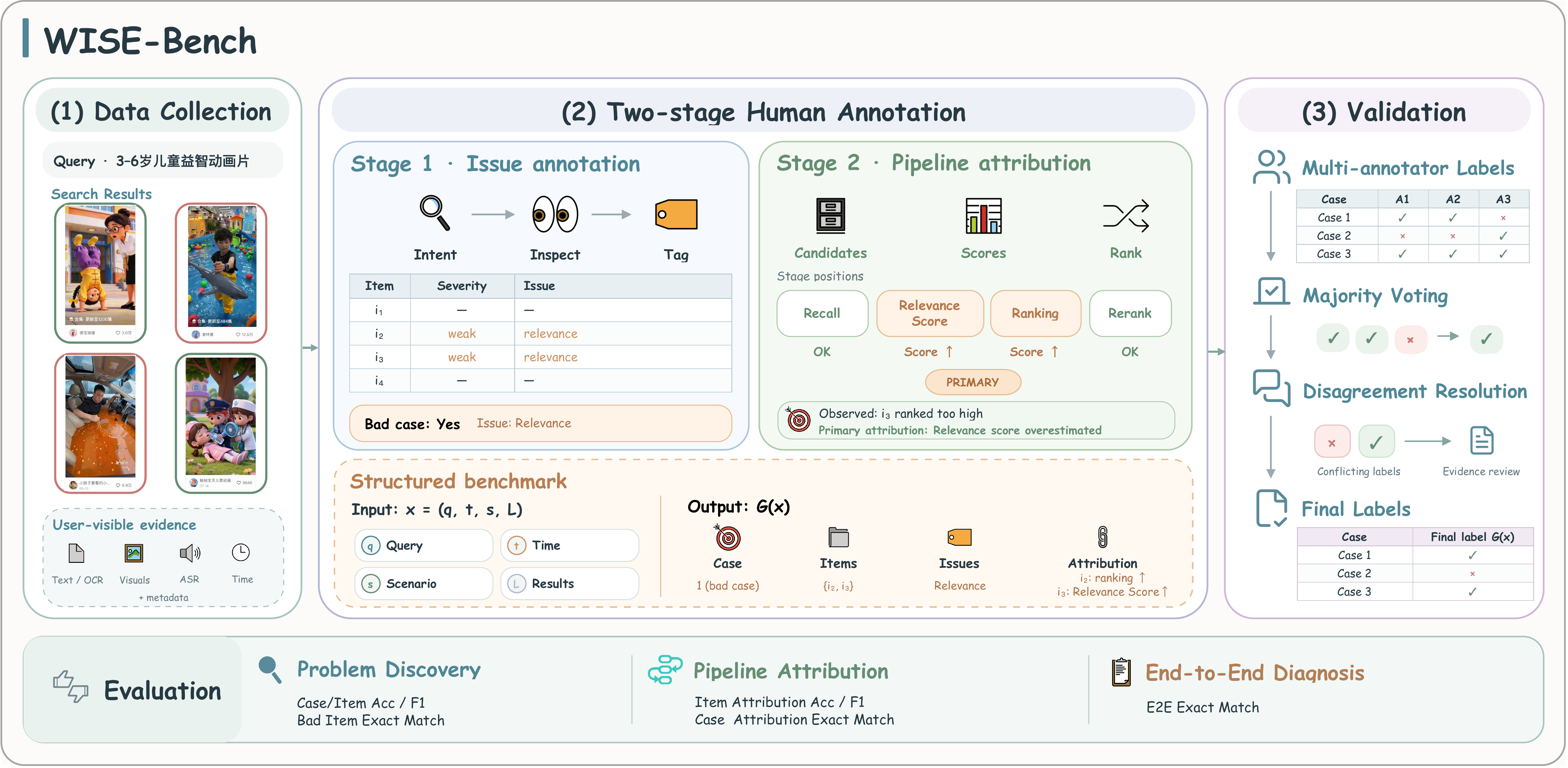}
    \caption{Construction and evaluation of \bench. Experts first label experience issues from request and content evidence, then inspect internal traces to annotate pipeline problems. The resulting labels support discovery, attribution, and end-to-end evaluation.}
    \label{fig:benchmark}
\end{figure*}

\subsection{WISE-BENCH}

Existing search benchmarks focus on result quality but lack internal evidence for pipeline attribution \cite{wei2024uniir,tao2025mmsearchplus,mao2026svfsearch}, while RCA benchmarks primarily target operational failures rather than search quality issues \cite{xu2025openrca,fang2026openrca2,gong2026orcabench,chen2025aiopslab}. To bridge this gap, we construct WISE-Bench with 12,000 real-world search cases, integrating multimodal experience evidence, pipeline traces, and expert annotations to support problem discovery, pipeline attribution, and end-to-end diagnosis.

\subsubsection{Data Selection}

WISE-Bench is constructed from Kuaishou online search requests. The requests are randomly sampled from the main search scenario of a large-scale short-video platform between July 1 and July 10. Each case contains the query, request time, search scenario, and displayed top-10 results. We collect two types of evidence. Experience evidence includes text, OCR, ASR, visual content, publication time, and author information. Pipeline evidence includes candidate supply, recall channels, model scores, and stage-wise ranking positions. We exclude records with missing or corrupted metadata, incomplete user-facing content, or inconsistent identifiers that prevent reliable linkage to pipeline traces. Queries that are clearly nonsensical or contain no meaningful search intent are excluded during data filtering to ensure reliable quality assessment.

\subsubsection{Data Annotation}
As illustrated in Figure~\ref{fig:benchmark}, WISE-Bench adopts a two-stage annotation protocol that separates experience assessment from pipeline attribution. In the first stage, annotators use only user-facing evidence to identify query-, case-, and item-level issues, including query ambiguity, page redundancy, problematic items, and their experience issues. In the second stage, annotators inspect internal pipeline traces to attribute confirmed problems. System-level failures, such as insufficient candidate supply or recall failures, are attributed at the case level without further item-level attribution. Otherwise, problematic items are attributed to relevance scoring, ranking, or reranking based on execution evidence. Annotations are finalized through majority voting and disagreement resolution. WISE-Bench contains 12,000 annotated cases, split into 8,000 training, 2,000 validation, and 2,000 test cases. Detailed annotation criteria and examples are provided in Appendix~\ref{app:a1}.

\subsubsection{Evaluation Tasks}

WISE-Bench contains multi-labels for experience issues and pipeline attribution. Since standalone page-level diversity issues are rare in the sampled traffic, our evaluation primarily focuses on case-level problem detection, item-level localization, and pipeline attribution. To systematically assess these capabilities, we define three core tasks: problem discovery, pipeline attribution, and end-to-end diagnosis.

\noindent\textbf{Problem discovery.} Given a search case and its experience evidence, the system predicts whether the result page contains an experience problem and identifies the problematic items. We report Case Accuracy over all search cases, including both normal and bad cases, and Item Accuracy over all displayed items. We additionally report Bad-item Exact Match, which measures the proportion of gold bad cases for which the predicted problematic-item set exactly matches the ground truth, with neither missing nor falsely identified items.

\noindent\textbf{Pipeline attribution.} Given gold problematic items and their pipeline evidence, the system predicts the responsible pipeline components. Providing gold items isolates attribution capability from upstream discovery errors. 
We report Item Attribution Accuracy over all gold problematic items with evaluable attribution labels. We also report Case Attribution Exact Match over cases containing at least one such item; a case is counted as correct only if every evaluated problematic item within it is correctly attributed.

\begin{figure*}[t]
    \centering
    \includegraphics[width=\textwidth]{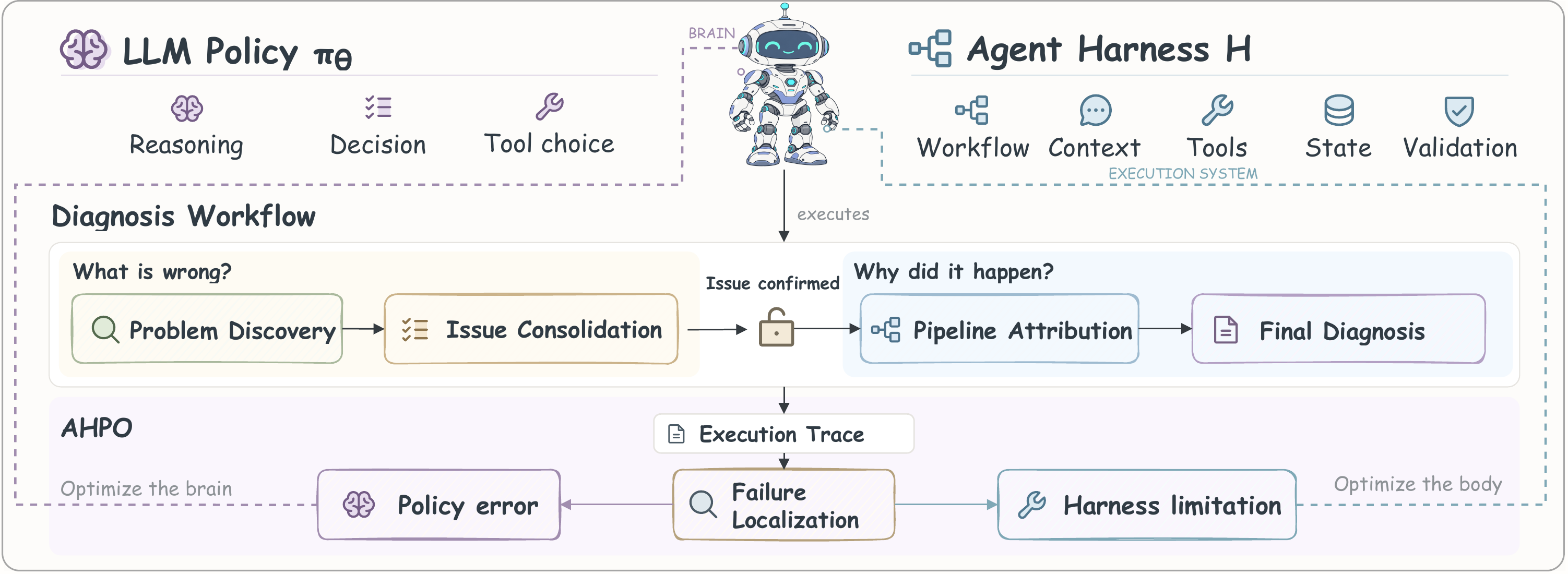}
    \caption{Overview of the \method framework.
    The bottom row shows the staged analysis process, while the agent harness
    coordinates model inference, context construction, evidence access,
    state management, and validation throughout execution.}
    \label{fig:fig1}
\end{figure*}

\noindent\textbf{End-to-end diagnosis.} The system receives the original search case and performs the complete diagnosis without gold problematic items. Let $\mathcal{T}$ denote all test cases, including normal and bad cases. Let $\delta_x$ and $\hat{\delta}_x$ denote the gold and predicted problem-presence indicators, and $I_x$ and $\hat{I}_x$ the gold and predicted problematic-item sets, respectively. Let $a_x$ indicate whether all case-level and item-level attributions match the gold annotations, excluding gold-unresolved fields from this comparison. We compute
\begin{equation}
\mathrm{E2E\ EM}
=
\frac{100}{|\mathcal{T}|}
\sum_{x\in\mathcal{T}}
\mathbf{1}\!\left[
\hat{\delta}_x=\delta_x
\land \hat{I}_x=I_x
\land a_x=1
\right].
\end{equation}
A case is correct only if its problem-presence label, problematic-item set, case-level supply/recall attributions, and all gold-item attributions are correct. For normal cases, this requires predicting no problem, an empty problematic-item set, and no case-level attribution. E2E EM focuses on problem localization and component attribution. Fine-grained issue labels $U$ are excluded to evaluate this core objective separately from issue categorization. Supporting evidence $E$ is excluded because different valid evidence sets may support the same attribution, making exact match against a single reference unsuitable for assessing evidence quality.

\section{WISE Framework}

\subsection{Agentic Formulation}

Search experience diagnosis is not a single-step prediction problem. It requires the system to first identify user-facing experience issues, then acquire the necessary internal evidence and attribute confirmed issues to specific stages of the search pipeline. These decisions depend on different sources of evidence and may involve intermediate reasoning and tool interactions, making the task naturally suited to an agentic formulation. Based on this formulation, we propose WISE, a diagnostic agent represented as
\begin{equation}
\mathcal{A}=(H,\pi_\theta),
\end{equation}
where $\pi_\theta$ is the LLM policy and $H$ is the agent harness. 
As illustrated in Figure~\ref{fig:fig1}, $\pi_\theta$ reasons, diagnoses, and selects actions, while $H$ manages context, available evidence, tools, state, and validation.

The bottom of Figure~\ref{fig:fig1} shows how the agent executes search experience diagnosis. WISE first discovers user-facing problems, consolidates the identified issues, and then performs pipeline attribution for confirmed issues before producing the final diagnosis. A key constraint throughout this process is the separation of evidence: problem discovery is performed using experience evidence only, while internal pipeline evidence is introduced only after an issue has been confirmed for attribution. The harness enforces this boundary by controlling the context and evidence available to the policy at each stage.

The same decomposition also provides the basis for optimizing WISE. As shown in Figure~\ref{fig:fig1}, a failed execution produces a diagnostic trace that is localized to the earliest supported failure. If the failure results from an incorrect decision under sufficient execution support, the policy is optimized; if the harness fails to provide the required evidence, action, or execution condition, the harness is revised. We detail the failure-driven optimization procedure in Section~\ref{sec:ahpo} and the agent workflow and harness implementation in Appendix~\ref{app:wise_details}.

\subsection{Policy Supervised Fine-Tuning}

Expert annotations provide the desired diagnostic outcomes but do not contain the execution trajectories needed for agent policy training. To bridge this gap, we first employ a teacher model to generate agent execution trajectories by performing the complete diagnostic workflow, including intermediate decisions, tool calls, and evidence acquisition. However, these initial trajectories may contain incorrect decisions or deviate from the expert annotations. Directly using such trajectories for supervised fine-tuning may propagate diagnostic errors, while correcting only their final outputs can introduce inconsistencies with the evidence observed during execution. We therefore introduce ReTrace to construct evidence-consistent supervision through minimal trajectory repair.

Formally, an execution trajectory is represented as $\tau=(x,d_1,a_1,\\ o_1,\ldots,d_T,y)$, where $x$ is the search case, $d_t$ denotes a structured decision, $a_t$ a tool call, $o_t$ the corresponding observation, and $y$ the final diagnostic output. Let $g$ denote the corresponding expert annotation. For trajectories inconsistent with $g$, ReTrace identifies the earliest repairable decision $t^*$ and divides the trajectory into an immutable prefix and a repairable suffix. All decisions, tool calls, and observations preceding $t^*$ are preserved. Starting from $t^*$, the teacher repairs the affected decisions and their downstream dependencies. When additional evidence is required, the teacher may introduce new tool calls, whose observations must be obtained through actual execution rather than generated or modified. The repaired trajectory is accepted only if its final diagnosis agrees with $g$ and its decisions are supported by the evidence available at the corresponding execution steps.

The accepted repaired trajectories are then used to construct supervised training data for the student policy. For each corrected decision node, we extract a training pair $(h,z')$, where $h$ is the execution prefix available at that node and $z'$ is the teacher-repaired response. The expert annotations serve as constraints for the teacher during repair but are excluded from the student's input. The student policy $\pi_\theta$ is initialized through autoregressive supervised fine-tuning:
\begin{equation}
\theta_{\mathrm{SFT}}
=
\arg\min_\theta
\sum_{(h,z')\in\mathcal{D}_{\mathrm{SFT}}}
-\log \pi_\theta(z'\mid h).
\end{equation}
This enables the student to learn diagnostic decisions grounded in actual execution evidence rather than merely imitating corrected final labels.

\subsection{Adaptive Harness--Policy Optimization}
\label{sec:ahpo}

After supervised fine-tuning, the diagnostic agent may still fail for different reasons. Some failures arise because the harness does not expose the required evidence, support the necessary action, or preserve the correct state. Others occur even when the execution environment is sufficient, because the policy misinterprets the available evidence or makes an incorrect decision. We therefore propose Adaptive Harness--Policy Optimization (AHPO), a failure-driven procedure that determines which optimization surface should be updated for each localized failure.

Starting from the current agent $A_k=(H_k,\pi_{\theta_k})$, we execute the complete diagnostic workflow on development cases and compare the resulting predictions with the benchmark annotations. For each failed case, a failure analyzer inspects the recorded execution trace and identifies the earliest observable decision that is inconsistent with the expected behavior and supported by the available evidence. The analysis records the erroneous node, the actual and expected decisions, the evidence available at that point, and the downstream outputs affected by the error. Missing evidence and suspected annotation problems are flagged separately rather than being forced into a model failure. The resulting localization is a trace-based account of error propagation, rather than a counterfactual causal claim.

Given a localized failure $f$, the intervention router determines whether the next update should target the harness or the policy. We consider three conditions: whether the evidence required for the correct decision was accessible, whether the harness supported the required action, and whether the current state representation and schema could express and preserve the correct result. Let $E(f)$, $A(f)$, and $S(f)$ denote these three conditions. The routing rule is
\begin{equation}
r(f)=
\begin{cases}
\text{HARNESS}, & \neg E(f)\lor\neg A(f)\lor\neg S(f),\\
\text{POLICY}, & E(f)\land A(f)\land S(f).
\end{cases}
\end{equation}
A HARNESS route indicates that the current execution framework prevents or obscures the correct decision. Such failures are grouped by shared mechanisms and converted into localized revisions of prompts, context construction, tool interfaces or boundaries, schemas, and validation or recovery rules. A POLICY route indicates that the required evidence and actions were available but the model nevertheless made an incorrect decision. These failures are collected as hard cases for the corresponding decision node and are combined with contrast cases near the decision boundary and anchor cases that the current policy already handles correctly. 
We use GRPO~\citep{shao2024deepseekmath} for policy-routed updates while keeping the harness fixed. We adopt a schema-gated hierarchical reward with node-specific evaluation criteria:
\begin{equation}
R_v(z,g)=
\begin{cases}
-\lambda_f, & \neg \operatorname{Valid}_v(z),\\
C_v(z,g)(1+\lambda_d D_v(z,g)), & \text{otherwise},
\end{cases}
\end{equation}
where $v$ denotes the target diagnostic node, $\operatorname{Valid}_v(z)$ checks whether the output satisfies the required schema, $C_v(z,g)\in\{0,1\}$ measures the correctness of the primary diagnostic decision, and $D_v(z,g)\in[0,1]$ evaluates dependent fine-grained decisions. The primary decision receives a base reward of 1 when correct, while $\lambda_d$ controls the contribution of fine-grained correctness and $\lambda_f$ penalizes invalid outputs. The reward prioritizes primary decision correctness, granting fine-grained rewards only when the primary decision is correct. The evaluation criteria are adapted to the diagnostic task at each node.

Each intervention produces a candidate agent by updating either the harness with the policy fixed or the policy with the harness fixed.
The candidate is re-evaluated through the complete diagnostic workflow using training and validation cases only; the held-out test set does not influence candidate acceptance. Updates are accepted only if they fix more cases than they regress, preserve all protected cases, and do not decrease case- or item-level accuracy on non-targeted evaluations. Rejected harness patches are reverted.
The accepted agent becomes $A_{k+1}$, after which the remaining failures are re-executed and localized again. In this way, harness revisions change the execution conditions under which subsequent policy decisions are made, while policy updates change the failure distribution that guides later harness revisions.

\begin{figure}[t]
    \centering
    \includegraphics[width=\linewidth]{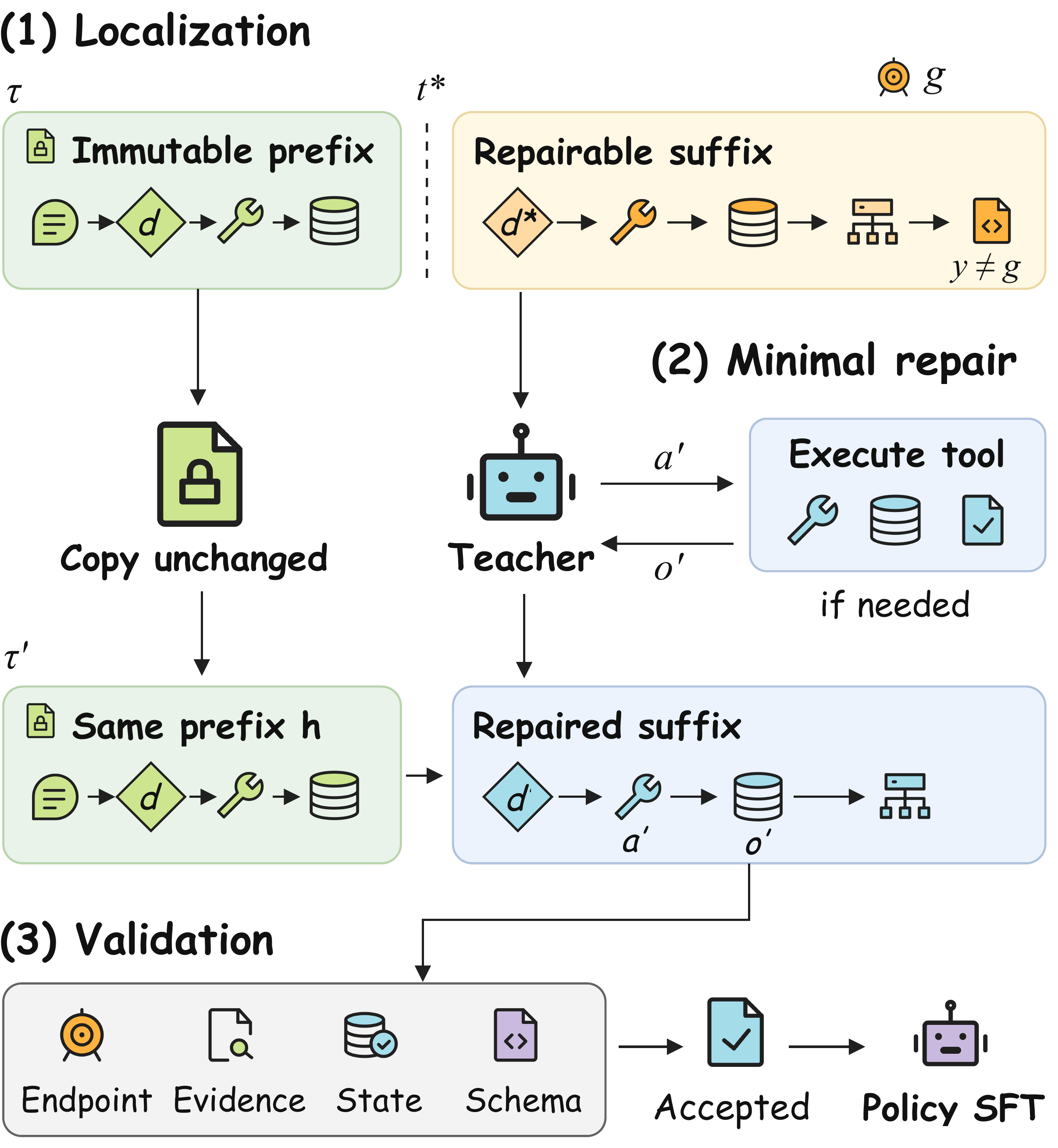}
    \caption{Overview of ReTrace. It preserves the unaffected prefix and repairs only the affected trajectory suffix with evidence-consistent execution.}
    \label{fig:repair}
\end{figure}

\begin{table*}[t]
    \centering
    \caption{Overall performance on \bench (\%).
    SFT uses ReTrace-constructed supervision.
    AHPO adaptively optimizes the harness and policy,
    while Harness Opt. updates only the harness with
    model weights fixed.}
    \label{tab:main}

    \small
    \setlength{\tabcolsep}{4.2pt}
    \resizebox{\textwidth}{!}{\begin{tabular}{llrrrrrr}
    \toprule
    \textbf{Model} & \textbf{Method} & \textbf{Case Acc.} & \textbf{Item Acc.} &
    \textbf{Bad-item EM} & \textbf{Item Attr. Acc.} & \textbf{Case Attr. EM} & \textbf{E2E EM} \\
    \midrule

    \rowcolor{black!6}
    \multicolumn{8}{l}{
        \textit{Harness optimization (fixed model weights)}
    } \\

    Kimi-K2.6 & Baseline
        & 67.95 & 72.29 & 44.86 & 25.63 & 16.23 & 48.55 \\
    & Harness Opt.
        & 73.05 & 79.84 & 49.57 & 30.15 & 23.67
        & 51.60 \\

    \addlinespace[1pt]

    GPT-5.6-Luna & Baseline
        & 64.25 & 69.95 & 39.41 & 41.82 & 20.32 & 50.10 \\
    & Harness Opt.
        & 67.30 & 73.23 & 41.64 & 47.36 & 26.77 & 52.15 \\

    \addlinespace[1pt]

    GPT-5.6-Terra & Baseline
        & 72.50 & 77.83 & 50.06 & 74.20 & 51.43 & 65.85 \\
    & Harness Opt.
        & 81.75 & 85.43 & 57.62 & 82.80 & 59.73
        & 69.75 \\

    \midrule

    \rowcolor{black!6}
    \multicolumn{8}{l}{
        \textit{SFT and Adaptive Harness--Policy Optimization}
    } \\

    Qwen3.5-4B & Baseline
        & 49.30 & 59.68 & 15.14 & 43.06 & 12.27 & 20.25\\
    & SFT
        & 73.00 & 83.91 & 51.92 & 86.67 & 57.87 & 58.25 \\
    & SFT + AHPO
        & 77.70 & 83.48 & 59.36 & 87.51 & 71.25
        & 69.85 \\

    \addlinespace[1pt]

    Qwen3.5-35B-A3B & Baseline
        & 64.20 & 71.41 & 29.74 & 59.87 & 15.86 & 35.25\\
    & SFT
        & 75.70 & 84.39 & 60.59 & 89.41 & 80.42 & 69.80 \\
    & SFT + AHPO
        & 80.25 & 84.48 & 64.19 & 89.95 & 81.29
        & 75.20 \\

    \bottomrule
    \end{tabular}
    }
\end{table*}

\section{Experiments}
\label{sec:experiments}

\subsection{Experimental Setup}

\subsubsection{Baselines}
We evaluate WISE on WISE-Bench using Qwen3.5-4B and Qwen3.5-35B-A3B~\cite{qwen35} under three settings: the base harness, supervised fine-tuning (SFT), and SFT followed by AHPO. AHPO uses GRPO~\cite{shao2024deepseekmath} for policy-routed updates while allowing harness-routed updates to revise the execution framework. WISE-Bench contains 12k annotated cases, split into 8k training cases, 2k validation cases, and 2k held-out test cases. To assess the benefits of harness optimization with fixed model weights, we additionally evaluate GPT-5.6-Luna, GPT-5.6-Terra~\cite{openai_gpt56}, and Kimi-K2.6~\cite{moonshot_kimi_k26} before and after harness optimization. Unless otherwise specified, reported results are evaluated on the held-out test set. 
Detailed model versions, decoding settings, training hyperparameters, and data splits are provided in Appendix~\ref{app:exp_details}.

\subsubsection{Metrics}
For evaluation, we report Case Accuracy (Case Acc.), Item Accuracy (Item Acc.), and Bad-item Exact Match (Bad-item EM) for problem discovery; Item Attribution Accuracy (Item Attr. Acc.) and Case Attribution Exact Match (Case Attr. EM) for pipeline attribution; and End-to-End Exact Match (E2E EM) for complete diagnosis. EM requires all relevant items in a case to be correctly predicted, thereby emphasizing complete case-level correctness over partial matches.

\subsection{Overall Performance}
\label{sec:overall}

Table~\ref{tab:main} reports the overall diagnostic performance on \bench.  For both Qwen backbones, SFT with ReTrace-constructed supervision substantially improves end-to-end diagnosis, and AHPO provides further gains. On Qwen3.5-4B, E2E EM increases from 20.25\% to 58.25\% after SFT and reaches 69.85\% with AHPO. On Qwen3.5-35B-A3B, it increases from 35.25\% to 69.80\% and then to 75.20\%. Taken together, these results show that the gains are not confined to the final prediction, but extend across multiple stages of the diagnostic workflow.

The improvements are particularly pronounced on case-level consistency metrics. For Qwen3.5-4B, AHPO improves Bad-item EM from 51.92\% to 59.36\% and Case Attribution EM from 57.87\% to 71.25\%, despite a slight decrease in Item Accuracy from 83.91\% to 83.48\%. For Qwen3.5-35B-A3B, Case Attribution EM improves more modestly, from 80.42\% to 81.29\%, while E2E EM increases by 5.40 percentage points. These metrics assess different aspects of diagnosis: attribution evaluation provides gold problematic items, whereas end-to-end evaluation requires the agent to discover and correctly attribute all problematic items. The results therefore highlight the importance of evaluating the complete workflow alongside its individual stages.

Table~\ref{tab:main} further shows that harness optimization provides consistent gains even when the underlying model weights are fixed. Harness Opt. improves all reported metrics for Kimi-K2.6, GPT-5.6-Luna, and GPT-5.6-Terra. For GPT-5.6-Terra, Case Accuracy increases from 72.50\% to 81.75\%, Case Attribution EM from 51.43\% to 59.73\%, and E2E EM from 65.85\% to 69.75\%. E2E EM also improves from 50.10\% to 52.15\% for GPT-5.6-Luna and from 48.55\% to 51.60\% for Kimi-K2.6. These gains further show that harness optimization can improve diagnosis without updating model weights.

\subsection{Effectiveness of ReTrace}
\label{sec:retrace_effectiveness}

Table~\ref{tab:trajectory} compares ReTrace with three alternative strategies for constructing trajectory supervision. Raw Trace retains the original execution without correction; Endpoint-only Correction modifies only the final target decision; and Full Regeneration regenerates the entire trajectory. In contrast, ReTrace preserves the unaffected prefix and repairs the affected suffix.

ReTrace achieves 57.87\% Case Attribution EM, outperforming Raw Trace (50.31\%) and Endpoint-only Correction (53.90\%), while matching the performance of Full Regeneration (57.87\%). It also achieves competitive performance on case-level and item-level accuracy. Compared with Full Regeneration, ReTrace reduces teacher-generation tokens from 32.0k to 20.4k, a 36.25\% reduction, without sacrificing Case Attribution EM. These results suggest that endpoint-only correction is insufficient to recover trajectory-level diagnostic performance, while full regeneration incurs substantially higher construction cost. ReTrace offers a favorable trade-off between diagnostic accuracy and trajectory construction efficiency.

\begin{table}[t]
    \centering
    \caption{Trajectory construction comparison. Accuracy and Attr. EM are reported in \%; Tokens measure teacher-generation cost.}
    \label{tab:trajectory}
    \small
    \setlength{\tabcolsep}{4pt}
    \begin{adjustbox}{max width=\linewidth}
\begin{tabular}{lrrrr}
    \toprule
    \textbf{Method} & \textbf{Case Acc.} & \textbf{Item Acc.} & \textbf{Case Attr. EM} & \textbf{Tokens} \\
    \midrule
    Raw
        & 71.25 & 82.88 & 50.31 & 0.0k \\
    Endpoint-only
        & \best{73.05} & 81.96 & 53.90 & 0.0k \\
    Full regen.
        & 71.75 & \best{84.62} & \best{57.87} & 32.0k \\
    \repair{} (ours)
        & 73.00 & 83.91 & \best{57.87} & 20.4k \\
    \bottomrule
    \end{tabular}
\end{adjustbox}
\end{table}

\subsection{Effectiveness of Adaptive Harness--Policy Optimization}
\label{sec:ahpo_effectiveness}

\begin{figure}[t]
    \centering
    \includegraphics[width=\linewidth]{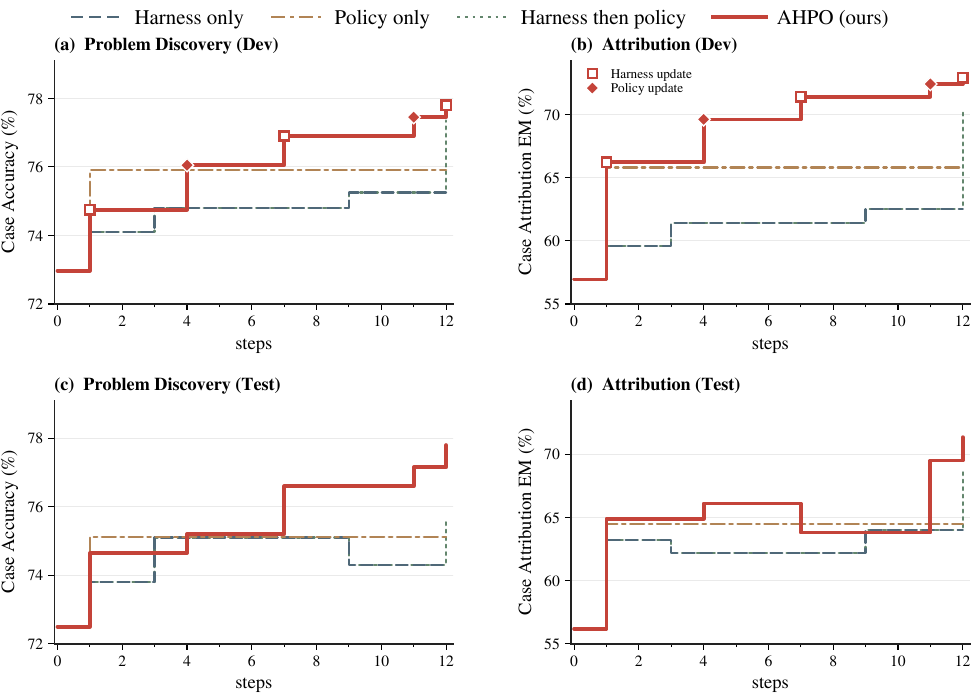}
    \caption{Development- and test-set performance across optimization iterations under different harness--policy update strategies. }
    \label{fig:evolution}
\end{figure}

We evaluate AHPO from two complementary perspectives: whether it can correctly identify the optimization surface associated with a diagnostic failure, and whether such failure-conditioned routing leads to more effective optimization than fixed update strategies. We first assess routing accuracy on manually adjudicated failures, then compare optimization strategies under matched intervention budgets.

We evaluate routing quality on item-level failure instances extracted from 200 manually adjudicated development cases, with each failure labeled as requiring either a harness or policy intervention. As shown in Table~\ref{tab:routing}, Direct Routing predicts the intervention target directly from the execution trace, achieving 58.9\% Accuracy and 43.6\% Macro-F1. Adding the Failure Analyzer, which localizes and summarizes the earliest supported failure, improves these scores to 69.2\% and 60.0\%, respectively. The full AHPO Router additionally applies the evidence-availability, action-support, and workflow-expressiveness checks defined in Section~\ref{sec:ahpo}, reaching 84.2\% Accuracy and 84.1\% Macro-F1. Its Harness F1 and Policy F1 are 83.5\% and 84.8\%, respectively, indicating comparable performance across the two intervention targets. These results show that reliable optimization routing benefits from both explicit failure localization and structured analysis of the execution conditions surrounding the failure.

We next compare AHPO with three alternative optimization strategies: Harness only, Policy only, and Harness then Policy, using the same number of candidate interventions and full-workflow evaluations. Figure~\ref{fig:evolution} separates the development-set performance used to guide optimization from the held-out test performance used to assess generalization. At each iteration, development-set failures guide candidate interventions, and development-set evaluation determines whether a candidate is accepted or rolled back. The test set is used only to evaluate the resulting agent versions and does not influence optimization or update acceptance. The development curves therefore show progress under the optimization criterion, whereas the test curves reveal whether these gains transfer to unseen cases. Harness only and Policy only show limited gains in later iterations, while Harness then Policy follows a fixed update schedule. AHPO achieves the strongest final performance on both sets, although intermediate development-set gains do not always translate into test-set improvements.

\begin{table}[t]
    \centering
    \caption{Failure-routing performance of different routing strategies.}
    \label{tab:routing}
    \small
    \setlength{\tabcolsep}{4.5pt}
    \begin{adjustbox}{max width=\linewidth}
\begin{tabular}{lcccc}
    \toprule
    \textbf{Router} &
    \textbf{Acc.} &
    \textbf{Macro-F1} &
    \textbf{Harness F1} &
    \textbf{Policy F1} \\
    \midrule
    Direct Routing
        & 58.9 & 43.6 & 14.2 & 73.0 \\
    + Failure Analyzer
        & 69.2 & 60.0 & 40.8 & 79.2 \\
    \textbf{AHPO Router}
        & \textbf{84.2}
        & \textbf{84.1}
        & \textbf{83.5}
        & \textbf{84.8} \\
    \bottomrule
    \end{tabular}
\end{adjustbox}
\end{table}

\section{Deployment and Applications}
\label{sec:online}

To evaluate the practical utility of WISE in a production search system, we integrated it into the search quality management workflow of a large-scale short-video platform. In the following subsections, we describe three representative use cases in which WISE has been applied in practice.

\subsection{Automated Quality Auditing}
\label{sec:automated_auditing}

Traditional search quality auditing relies on periodic manual evaluations and bad-case feedback. Human evaluators first identify search experience issues, after which engineers inspect internal pipeline signals to investigate their causes. This process is labor-intensive and difficult to scale. WISE automates this process by identifying experience issues from search results and tracing them to underlying pipeline components, replacing much of the manual evaluation and investigation effort. Table~\ref{tab:diagnosis_efficiency} compares the efficiency and cost of WISE with the traditional manual workflow. With concurrent inference, WISE processes approximately 120K search cases per day, with an estimated cost of RMB 25.60 per 1,000 cases, compared with RMB 5,714.29 for manual diagnosis. This represents an estimated cost reduction of 99.55\%, highlighting the efficiency of automated diagnosis at production scale.

\begin{table}[t]
\centering
\caption{Operational efficiency and estimated diagnosis cost
of manual auditing and WISE. Average latency in s/case; throughput in cases/day; cost in RMB/1K cases.}
\label{tab:diagnosis_efficiency}
\small
\begin{adjustbox}{max width=\linewidth}
\begin{tabular}{lcc}
\toprule
\textbf{Metric} & \textbf{Manual} & \textbf{WISE} \\
\midrule
Discovery Time    & 180.00 s & 23.9 s\\
Attribution Time  & 300.00 s & 37.6 s\\
End-to-End Time   & 480.00 s & 61.5 s\\
Throughput        & 52.5/person & 120K/cluster \\
Cost              & 5,714.29 & 25.60 \\
\bottomrule
\end{tabular}
\end{adjustbox}
\end{table}

\subsection{Online Quality Assurance}
Frequent model updates and A/B experiments in industrial search systems may introduce localized quality regressions that are not reflected in aggregate business metrics. To address this, we deploy WISE for online quality assurance. We use the optimized Qwen3.5-35B-A3B as the backbone, with eight inference instances, each equipped with four NVIDIA L20 GPUs. WISE performs daily inspections of approximately 120,000 search requests across different experiment buckets, identifying search experience issues and attributing them to pipeline components.

By aggregating diagnosis results for each bucket, we monitor the incidence of different quality issues and detect abnormal changes through statistical comparisons across buckets and analysis of historical trends. This enables us to identify and quantify fine-grained quality regressions following model or strategy rollouts, even when aggregate business metrics remain stable. Pipeline attribution further reveals how these issues arise within the search pipeline, providing insights for evaluating and improving deployed strategies.

\subsection{WISE-Guided Model Optimization}

Beyond online quality assurance, WISE's diagnostic results can further guide the optimization of different components in the search pipeline. For example, identified intent understanding errors can inform intent model improvements, while retrieval failures can provide targeted examples for optimizing recall strategies. As an initial case study, we focus on relevance model optimization. Specifically, we leverage WISE's pipeline attribution to identify relevance-overestimation cases, where the relevance model assigns high scores to results that fail to satisfy the user's search intent. Over a 30-day period, we collect approximately 2.2 million such query--item samples and incorporate them into the relevance model's training data to improve its ability to distinguish relevant from superficially relevant results.

\subsubsection{Offline Evaluation}

We compare the baseline relevance model with WISE-Guided Training, which augments the original training data with WISE-mined samples. Both models share the same architecture and training configuration and are evaluated on 3,385 held-out query--item pairs using MSE, Accuracy, Precision, Recall, and F1. As shown in Table~\ref{tab:relevance_optimization}, WISE-Guided Training improves all reported metrics. MSE decreases from 0.0638 to 0.0606 (5.0\% relative reduction), while F1 increases from 0.573 to 0.631 (10.2\% relative improvement), demonstrating improved relevance prediction.

\subsubsection{Online Evaluation}

We further evaluate the WISE-guided relevance model in the Kuaishou production search system. In a human GSB evaluation comparing the top-$4$ search results against the production baseline, WISE-Guided Training achieves a 4.16\% advantage, as shown in Table~\ref{tab:online_gsb}. We also conduct a 7-day online A/B test covering 5\% of search traffic, measuring Intent Query Reformulation Rate (IQRR), Click-Through Rate (CTR), and Long-View Ratio (LVR). Compared with the production baseline, WISE-Guided Training reduces IQRR by 0.11\% (CI: [-0.17\%, -0.05\%]), while increasing CTR by 0.15\% (CI: [0.06\%, 0.24\%]) and LVR by 0.20\% (CI: [0.07\%, 0.33\%]). These results are consistent with the improvements observed in offline evaluation and human assessment.

\begin{table}[t]
    \centering
    \caption{Relevance model performance with and without WISE-guided training, reported as weighted averages across test sets for three relevance types.}
    \label{tab:relevance_optimization}
    \setlength{\tabcolsep}{6pt}
    \begin{adjustbox}{max width=\linewidth}
\begin{tabular}{lccccc}
        \toprule
        \textbf{Method}
        & \textbf{MSE} $\downarrow$
        & \textbf{Acc.} $\uparrow$
        & \textbf{P.} $\uparrow$
        & \textbf{R.} $\uparrow$
        & \textbf{F1} $\uparrow$ \\
        \midrule
        Baseline
        & 0.0638 & 0.612 & 0.503 & 0.665 & 0.573 \\
        + WISE-Guided
        & \textbf{0.0606}
        & \textbf{0.636}
        & \textbf{0.558}
        & \textbf{0.727}
        & \textbf{0.631} \\
        \bottomrule
        
    \end{tabular}
\end{adjustbox}
\end{table}

\begin{table}[t]
\centering
\caption{Human GSB evaluation against the production
baseline. Annotators compare the top-$K$ results
of each method with the baseline and label
Good/Same/Bad.
Adv $= (G-B)/(G+S+B)\times 100\%$.}
\label{tab:online_gsb}
\small
\begin{adjustbox}{max width=\linewidth}
\begin{tabular}{lccccc}
\toprule
\textbf{Method} & \textbf{Good} & \textbf{Same} &
\textbf{Bad} & \textbf{Adv} & \textbf{$p$-value} \\
\midrule
WISE-Guided & 387 & 1304 & 304 & 4.16\% & $<0.01$ \\
\bottomrule
\end{tabular}
\end{adjustbox}
\end{table}

\section{Conclusion}

In this paper, we formulate System Quality Diagnosis, extending conventional quality evaluation beyond identifying poor outcomes to attributing their causes across underlying system components. Focusing on industrial search, we propose WISE, an agentic framework that integrates multimodal experience assessment with evidence-grounded pipeline attribution, and construct \bench, a real-world benchmark combining user-facing evidence, internal pipeline traces, and expert annotations. To improve diagnostic reliability, we introduce ReTrace for evidence-consistent trajectory supervision and AHPO for failure-guided harness--policy optimization.
Experiments show that WISE improves end-to-end diagnosis success from 35.25\% to 75.20\% on Qwen3.5-35B-A3B. 
Beyond offline evaluation, WISE has been deployed in a large-scale production search system, supporting a broad range of applications, including automated quality auditing, online quality assurance, and relevance model optimization. 
These results demonstrate the practical value of integrating quality evaluation, pipeline attribution, and downstream optimization in industrial search.

\bibliography{refs}
\bibliographystyle{plainnat}

\clearpage
\appendix
\section{Annotation Details}
\label{app:a1}

\subsection{Quality Control}

The 12,000 cases are randomly shuffled and distributed among 14 annotators. Each case is independently labeled by two annotators without access to each other's judgments. Disagreements are randomly assigned to a third annotator who did not participate in the initial labeling, and the final label follows the majority decision. When all three annotators assign different component labels, a reviewer examines the evidence and determines the final attribution. Table~\ref{tab:annotation-agreement} reports percentage agreement and nominal Krippendorff's $\alpha$ for case-level and item-level problem identification, and primary component attribution on jointly confirmed problems. All statistics use the initial two independent annotations, excluding third-annotator judgments and adjudicated labels.

\begin{table}[h]
    \centering
    \caption{Inter-annotator agreement.}
    \label{tab:annotation-agreement}
    \small
    \begin{tabular*}{\columnwidth}{@{\extracolsep{\fill}}lrr@{}}
        \toprule
        \textbf{Annotation} & \textbf{Agreement} & \textbf{$\alpha$} \\
        \midrule
        Bad-case decision & 93.1\% & 0.856 \\
        Bad-item decision & 89.7\% & 0.678 \\
        Component attribution & 94.3\% & 0.915 \\
        \bottomrule
    \end{tabular*}
\end{table}

\subsection{Annotation Details}

\noindent\textbf{Experience issues.}
Experience issues include \textit{relevance}, \textit{freshness}, \textit{authority}, \textit{content quality}, and \textit{diversity}. All except diversity are annotated at the item level: relevance assesses whether the content satisfies the query intent; freshness assesses whether it meets temporal requirements; authority assesses whether the source meets the query's credibility requirements; and content quality assesses whether visual, audio, or substantive defects impair viewing or information acquisition. Diversity is annotated at the case level, identifying excessive repetition or insufficient coverage of query intent. An item may receive multiple issue labels, and a case may contain both item-level issues and diversity issues. Table~\ref{tab:annotation-example} illustrates this distinction: Item C has a relevance issue due to a species mismatch, while no content-quality or case-level diversity issues are identified.

\smallskip
\noindent\textbf{Pipeline attribution.}
Annotators use candidate supply, recall paths, model scores, and stage-wise ranking positions to attribute confirmed issues to \textit{supply}, \textit{recall}, \textit{relevance scoring}, \textit{ranking}, or \textit{reranking}. Supply limitations occur when the corpus lacks suitable content; recall failures occur when suitable content exists but is absent from the retrieved candidates. Relevance-scoring errors underestimate relevant content or overestimate irrelevant content. Ranking errors involve inappropriate ordering that cannot be explained solely by upstream relevance-scoring errors. Reranking errors arise when subsequent adjustments introduce or aggravate experience issues, such as promoting irrelevant items or reducing diversity. Attribution requires internal execution evidence rather than the displayed results alone. For example, Item C in Table~\ref{tab:annotation-example} receives a relevance score of 0.819 despite the species mismatch, with no position change during reranking; relevance scoring is therefore selected as the primary attribution. When the evidence is insufficient to distinguish a primary component, attribution is marked as \textit{unresolved}, while the confirmed experience issue is retained.

\smallskip
\noindent\textbf{Final labels.}
Annotations are retained at two levels:
\begin{itemize}
    \item \textbf{Case level:} whether the page contains an experience issue, diversity labels, and overall attributions such as supply limitations or recall failures.
    \item \textbf{Item level:} the problematic item's identifier and displayed position, experience-issue labels, and one primary component among relevance scoring, ranking, and reranking.
\end{itemize}
Since upstream errors may propagate to later stages, downstream consequences are not separately labeled as independent causes. The primary component is determined from evidence across stages, and supporting attribution evidence is retained at both levels.

\begin{table}[t]
    \centering
    \caption{A WISE-Bench example with experience labels and pipeline attribution.}
    \label{tab:annotation-example}

    \begingroup
    \footnotesize
    \setlength{\tabcolsep}{3pt}
    \renewcommand{\arraystretch}{1.12}

    \begin{tabular}{@{}
        p{0.08\columnwidth}
        p{\dimexpr0.67\columnwidth-4\tabcolsep\relax}
        p{0.25\columnwidth}
    @{}}
        \toprule
        \multicolumn{3}{@{}l}{\textbf{Query: Yellow catfish}} \\
        \midrule
        \textbf{Item} & \textbf{Video content} & \textbf{Item-level issue} \\
        \midrule

        A &
        A prepared yellow-catfish dish. &
        None \\

        \addlinespace[2pt]

        B &
        A tutorial on making yellow-catfish soup. &
        None \\

        \addlinespace[2pt]

        C &
        Silver-gray, striped fish in a container; the title names another species. &
        Relevance \\

        \addlinespace[2pt]

        D &
        Cleaning and preparing yellow catfish before cooking. &
        None \\

        \addlinespace[2pt]

        E &
        A home-style braised yellow-catfish recipe. &
        None \\

        \bottomrule
    \end{tabular}

    \par\smallskip
    \noindent
    \begin{minipage}{\columnwidth}
        \footnotesize
        \raggedright
        \textbf{Experience judgment.}
        Items A, B, D, and E match the queried species. Item C concerns a different species and is marked as a relevance issue. No content-quality issues are identified. The results cover dish presentation, soup preparation, cleaning, and braising without excessive repetition; no case-level diversity issue is identified.
    \end{minipage}

    \par\medskip

    \begin{tabular}{@{}
        p{0.27\columnwidth}
        p{\dimexpr0.73\columnwidth-2\tabcolsep\relax}
    @{}}
        \toprule
        \multicolumn{2}{@{}l}{\textbf{Attribution evidence for Item C}} \\
        \midrule

        Retrieval &
        ANN channel. \\

        \addlinespace[2pt]

        Relevance &
        Score: 0.819; overestimation threshold: 0.650. \\

        \addlinespace[2pt]

        Ranking &
        Click prediction: 0.1046 (98.67th percentile); long-view prediction: 0.1167 (99.50th percentile). \\

        \addlinespace[2pt]

        Reranking &
        Position change: $\Delta r = 0$. \\

        \midrule

        \textbf{Primary component} &
        \textbf{Relevance scoring.} The favorable relevance score conflicts with the species mismatch. No further promotion is observed during reranking. \\

        \midrule

        \textbf{Final labels} &
        Case: bad experience; no diversity issue.\newline
        Problematic items: $\{C\}$.\newline
        Item C: relevance issue; primary attribution to relevance scoring.\newline
        Content-quality issues: none. \\

        \bottomrule
    \end{tabular}

    \endgroup
\end{table}

\section{WISE Workflow}
\label{app:wise_details}

\subsection{Multi-Agent Architecture}

WISE is implemented as a LangGraph-based hierarchical multi-agent workflow comprising a problem-discovery team, an attribution team, and a report-generation agent. Agent execution is organized into graph nodes, with conditional edges controlling task routing and stage transitions. \textsf{QueryCheck} interprets the query; \textsf{ProblemDiscoveryMaster} dispatches inspections to specialists; and \textsf{ProblemDiscoverySummary} consolidates their findings. \textsf{AttributionMaster} coordinates specialists for supply/recall, relevance scoring, ranking, and reranking. Each role has dedicated instructions and tool permissions. A separate, tool-free agent, \textsf{FinalSummary}, renders the diagnostic report.

\subsection{Communication and Validation}

Agents exchange structured JSON records through the harness, with LangGraph's shared state carrying information between nodes. Inputs contain case and task identifiers, the inspection scope, relevant item identifiers, upstream findings, and accessible evidence references. Outputs follow role-specific schemas specifying required fields, field types, and allowed diagnostic labels. The harness checks schema compliance, identifier consistency, and evidence references before forwarding outputs; these checks establish structural validity rather than diagnostic correctness. Invalid responses trigger bounded correction attempts, and persistent failures are recorded separately from negative judgments. The shared state maintains query interpretation, experience issues, specialist findings, consolidated attributions, and execution status for each case, preserving specialist findings separately from consolidated decisions. Each agent's input context is constructed from the state fields and evidence permitted for its role rather than exposing the entire state. Agent messages, tool observations, and validation results are retained in the execution trace for trajectory construction and failure analysis.

\subsection{Execution and Consolidation}

\textsf{QueryCheck} writes its intent representation to the shared state, which \textsf{ProblemDiscoveryMaster} uses to dispatch item-level and page-level inspections. \textsf{ProblemDiscoverySummary} consolidates the specialists' JSON findings into the input to \textsf{AttributionMaster}. Conditional routing sends cases without confirmed problems directly to final reporting, while incomplete inspections remain explicitly marked. \textsf{AttributionMaster} coordinates case-level supply/recall assessment and applicable item-level investigations. Specialist findings are consolidated into case-level attributions for supply and recall issues and item-level attributions identifying a primary component when sufficient evidence is available. Insufficient evidence leaves the attribution unresolved without removing the confirmed issue. \textsf{FinalSummary} renders the consolidated records without changing diagnostic labels or introducing new claims; the structured JSON remains the authoritative output for evaluation and downstream processing.

\section{Prompts for Trajectory Construction}
\label{app:trajectory-implementation}

We implement the four conditions in Table~\ref{tab:trajectory} as separate skills: \textsf{trajectory-raw} retains the original execution; \textsf{trajectory-endpoint-correction} modifies only the scored endpoint fields; \textsf{trajectory-teacher-regeneration} regenerates all assistant--tool interactions; and \textsf{trajectory-minimal-repair} freezes the prefix before the earliest repairable decision and minimally repairs the affected suffix. The following prompts specify the two teacher-based conditions, with deployment-specific values shown as placeholders.

\subsection{Full Teacher Regeneration}

Full Regeneration regenerates all assistant--tool interactions from the original system and user messages, obtaining tool observations through actual execution. The adjudicated endpoint serves only as a teacher-side target, not evidence.

\begin{lstlisting}[style=prompt,breaklines=true,breakatwhitespace=true,caption={Teacher prompt for full trajectory regeneration.},label={lst:teacher-regeneration}]
[SYSTEM MESSAGE]
Regenerate the complete assistant--tool trajectory from the original system and user messages. Do not reuse previously generated interactions.
Treat the adjudicated endpoint as a target, not evidence. Do not cite it as evidence.
Follow the diagnostic workflow and use only evidence available at each step.
Obtain all tool observations through actual execution. Never fabricate observations.
Follow the required output schemas throughout the workflow. The final diagnosis must match the adjudicated endpoint.
If available or obtainable evidence cannot support the target, report failure.
[USER MESSAGE]
<ORIGINAL_INPUT>
<ADJUDICATED_ENDPOINT>
<AVAILABLE_TOOLS_AND_SCHEMAS>
\end{lstlisting}

\subsection{Evidence-Consistent Minimal Repair}

For ReTrace, we identify the earliest repairable decision $t^\star$ and divide the recorded trajectory into an immutable prefix and a repairable suffix. All messages, decisions, tool calls, and observations preceding $t^\star$ remain unchanged. Given this frozen prefix, the original suffix, and the adjudicated endpoint as a target rather than evidence, the teacher minimally repairs the affected decision and its downstream dependencies. When additional evidence is needed, new tool calls are executed in the actual environment.

A repaired trajectory is accepted only if its final output matches the adjudicated endpoint, its diagnostic claims are supported by evidence available at the corresponding execution steps, and its tool interactions satisfy the workflow constraints. The repair must also preserve case and item identity and follow the required output schemas. Trajectories that fail these checks are rejected rather than forced to match the endpoint. For each accepted repair, we extract supervised examples $(h,z')$ from the corrected decision nodes, where $h$ is the execution prefix available at that node, excluding teacher-only endpoint annotations, and $z'$ is the repaired response.

\begin{lstlisting}[style=prompt,breaklines=true,breakatwhitespace=true,caption={Teacher prompt for evidence-consistent minimal trajectory repair.}]
[SYSTEM MESSAGE]
You are repairing a recorded agent execution trajectory.
Preserve all messages, decisions, tool calls, and observations before the repair point.
Treat the adjudicated endpoint as a target, not evidence. Do not cite it as evidence.
Starting at the repair point, minimally repair the affected decision and its downstream dependencies.
If additional evidence is needed, use workflow-authorized tools.
Obtain all new observations through actual tool execution. Never fabricate or alter tool observations.
Use only evidence available at each decision node. Follow workflow constraints and the required output schema for the target node.
If available or obtainable evidence cannot support the target, mark the trajectory as unrepairable.
[USER MESSAGE]
<FROZEN_PREFIX_BEFORE_REPAIR_POINT>
<ORIGINAL_SUFFIX>
<REPAIR_POINT>
<ADJUDICATED_ENDPOINT>
<AVAILABLE_TOOLS_AND_SCHEMAS>
\end{lstlisting}

\section{Skills for Harness--Policy Optimization}
\label{app:ahpo-implementation}

The following specifications define the four roles in AHPO and correspond to the configurations used in our reported experiments. The routing rules distinguish harness-side limitations from incorrect policy decisions. Repository-specific commands and deployment controls are omitted for clarity.

\subsection{Failure Analyzer}

\begin{lstlisting}[style=prompt,caption={Methodological specification for \texttt{search-failure-analyzer}.}]
Role: Analyze Search Diagnosis ProblemDiscovery trajectories against Gold annotations and locate the earliest supported decision error. Do not propose a harness change or assign a HARNESS/POLICY route.

Input:
- benchmark input and Gold annotation;
- model prediction and evaluator mismatch fields;
- compact execution trace and referenced full trajectory records.

Procedure:
1. Inspect the observable execution sequence in order: QueryCheck, Master routing, evidence collection, sub-agent execution, sub-agent judgment, and merge/summary.
2. Determine whether the mismatch is caused by a model decision, an upstream tool/data/environment failure, an annotation problem, or insufficient evidence.
3. For a model failure, identify the earliest observable erroneous decision that plausibly explains the scored mismatch. Mark this as a trace-based hypothesis unless intervention and replay have verified it.
4. Record the actual behavior and the expected constraint at that node.
5. Trace how the earliest error propagated into later decisions and the final prediction.
6. Record correct upstream decisions as protected behavior that later optimization should preserve.

Use only observable trajectory evidence; do not infer hidden reasoning. If required evidence is absent, return INSUFFICIENT_EVIDENCE. If the Gold annotation appears wrong, return ANNOTATION_SUSPECTED rather than manufacturing a model error.

Output one structured record with:
- verdict and failure domain;
- scored mismatch;
- earliest_error {stage, node, agent, decision_type,
  actual_behavior, expected_constraint, evidence};
- causal_chain and reusable failure_mechanisms;
- available_evidence and missing_evidence;
- protected_behaviors, confidence, and needs_human_review.
\end{lstlisting}

\subsection{Intervention Router}

\begin{lstlisting}[style=prompt,caption={Methodological specification for \texttt{search-intervention-router}.}]
Role: Route a validated Search Diagnosis failure to the harness or the model policy. Determine the optimization surface; do not propose the concrete fix or training examples.

Input: one validated failure analysis containing its earliest error, causal chain, available and missing evidence, and protected behavior.

Answer the following questions in order using observable evidence from before the first erroneous decision:
1. Evidence availability: Was the evidence required for the correct decision available under the current environment and tool boundary?
2. Action support: Did the harness permit the required action and correctly execute actions that the policy requested, including routing, tool use, validation, and recovery? Distinguish an unavailable or incorrectly executed action from the policy choosing the wrong available action.
3. Workflow expressiveness: Could the current schema and workflow express and preserve the correct result?

Routing rule:
- If any answer is no because of a harness constraint or execution defect, route to HARNESS. Identify the workflow, prompt, tool, context, schema, validation, or recovery condition requiring revision.
- If all three answers are yes but the node makes an incorrect decision or selects the wrong available action, route to POLICY. State the remaining capability gap, such as intent understanding, evidence interpretation, attribution, abstention, or action selection.
- Treat the route as the next intervention to evaluate, not an exclusive assignment of responsibility. A prompt change and a policy update may both help the same case.
- If a question cannot be answered from the available evidence, return no route and request further review.

Output one structured record with:
- evidence_available {answer, reason, evidence};
- action_supported_and_executed {answer, reason, evidence};
- schema_workflow_expressive {answer, reason, evidence};
- route, rationale, capability_gap, and confidence.
\end{lstlisting}

\subsection{Harness Optimizer}

\begin{lstlisting}[style=prompt,caption={Methodological specification for \texttt{search-harness-optimizer}.}]
Role: Convert failures routed to HARNESS into minimal, generalizable harness revisions.

Input:
- validated HARNESS routing results and their failure analyses;
- current workflow, prompts, tool definitions and boundaries,
  context construction, output schemas, validators, recovery logic,
  and relevant skill instructions;
- current results for failed and correctly solved cases.

Procedure:
1. Normalize each failure by target node, harness component, and causal mechanism.
2. Form a cluster only when one causal change can address all included failures, or when one case reveals a high-impact structural invariant.
3. Identify the shared root cause rather than copying case-specific symptoms.
4. Choose exactly one primary harness component: workflow, prompt, tool_definition, tool_boundary, context, schema, validation, recovery, or skill.
5. Propose the smallest shared change that makes the expected behavior explicit or mechanically enforceable.
6. Select target cases exhibiting the failure, nearby contrast cases around the decision boundary, and already-correct anchor cases.
7. State the invariant expected after the change and the behaviors most at risk of regression.

Output a structured proposal containing:
- cluster_id, target_node, and harness_component;
- normalized failure_mechanism and shared_root_cause;
- supporting failure records;
- proposed_change and expected_invariant;
- target, contrast, and anchor case sets;
- affected harness surfaces and regression_risks.
\end{lstlisting}

\subsection{Training Data Planner}

\begin{lstlisting}[style=prompt,caption={Methodological specification for \texttt{search-training-data-planner}.}]
Role: Convert failures routed to POLICY into capability-focused training groups for agentic policy optimization.

Input:
- validated POLICY routing results and failure analyses;
- benchmark examples and current-agent rollout results;
- the current harness and policy versions.

Procedure:
1. Group failures by target node and capability gap rather than by surface wording or query identity.
2. For each group, select hard cases backed by a POLICY route at that node.
3. Add contrast cases near the relevant decision boundary so that the target behavior is discriminative rather than a blanket rule.
4. Add anchor cases that the current policy already solves, preserving established behavior during optimization.
5. Target the failing node and its relevant decision records or sub-trajectories. Document whether model parameters are shared across nodes; verify the complete workflow after a targeted update.
6. Choose a reward type: rule-based for directly checkable behavior, outcome-based for endpoint correctness, or hybrid when both process and outcome matter.
7. Define the Gold-derived target behavior and scoring dimensions for each selected case.

Output a structured plan containing:
- capability clusters with target_node and capability_gap;
- each selected case and its hard/contrast/anchor role;
- training_scope and reward_type;
- the target behavior and scoring dimensions for optimization.
\end{lstlisting}

\section{Experimental Details}
\label{app:exp_details}

\subsection{Data Splits}

All splits are performed at the search-case level before node-level training records are extracted. Thus, node records originating from the same search case never appear across training, development, and test sets. The test set is excluded from trajectory construction, harness optimization, policy updates, and model selection.

\subsection{Training and Inference Configuration}

\begin{table}[h]
    \centering
    \caption{Distribution of node-level supervision used for SFT.}
    \label{tab:sft_data}
    \small
    \begin{tabular}{lrr}
    \toprule
    \textbf{Node Type} & \textbf{Unique Node Records} & \textbf{Fraction (\%)} \\
    \midrule
    Problem discovery     & 21,178 & 58.4 \\
    Issue consolidation   & 3,262  & 9.0 \\
    Pipeline attribution  & 9,660  & 26.6 \\
    Final diagnosis       & 2,180  & 6.0 \\
    \midrule
    \textbf{Total }                & 36,280 & 100.0 \\
    \bottomrule
    \end{tabular}
\end{table}

The evaluated backbones are Qwen3.5-4B, Qwen3.5-35B-A3B, GPT-5.6-Luna, GPT-5.6-Terra, and Kimi-K2.6. ReTrace uses GPT-5.6-Terra for trajectory construction. Unless otherwise specified, inference uses a temperature of 0.01 and a maximum generation length of 32,768 tokens. Comparisons under the same backbone use identical decoding settings.

SFT uses 36,280 node-level training records extracted from accepted ReTrace repairs, with their distribution across workflow stages shown in Table~\ref{tab:sft_data}. We train for 2 epochs with a learning rate of $1\times10^{-5}$ and a global batch size of 64, using 8 GPUs with a per-device batch size of 1 and gradient accumulation of 8. The maximum sequence length is 24,576 tokens. 
For policy-routed AHPO updates, GRPO uses 8 rollouts per training instance and a learning rate of $5\times10^{-7}$. For the hierarchical reward, we set $\lambda_f = 1 $ and $\lambda_d = 0.7$ across diagnostic nodes. For each policy-routed intervention, GRPO trains on the selected data for 1 epochs with a batch size of 16. The number of training steps varies with the dataset size.

\end{document}